\documentclass[12pt]{iopart}
\usepackage[dvips]{graphicx}
\begin{document}

\title{Searching for Perfect Fluids:\\ Quantum Viscosity in a Universal Fermi Gas}
\author{C. Cao, E. Elliott, H. Wu, J. E. Thomas}
\address{Physics Department, Duke University, Durham, North
Carolina 27708-0305} \pacs{03.75.Ss, 32.80.Pj}

\date{\today}

\begin{abstract}
We measure the shear viscosity in a two-component Fermi gas of atoms, tuned to a broad s-wave collisional (Feshbach) resonance. At resonance, the atoms strongly interact and exhibit universal behavior, where the equilibrium thermodynamic properties and the transport coefficients are universal functions of the density $n$ and temperature $T$. We present a new calibration of the temperature as a function of global energy, which is directly measured from the cloud profiles. Using the calibration, the trap-averaged shear viscosity in units of $\hbar\,n$ is determined as a function of the reduced temperature at the trap center, from nearly the ground state to the unitary two-body regime. Low temperature data is obtained from the damping rate of the radial breathing mode, while high temperature data is obtained from hydrodynamic expansion measurements.  We also show that the best fit to the high temperature expansion data is obtained for a vanishing bulk viscosity. The measured trap-averaged entropy per particle and shear viscosity are used to estimate the ratio of the shear viscosity to the entropy density, which is compared that conjectured for a perfect fluid.
\end{abstract}

\maketitle

\section{Introduction}
\label{sec:intro}
The measurement of the shear viscosity is currently of particular interest in the
context of a recent conjecture, derived using string theory methods,
which defines a perfect normal fluid~\cite{Kovtun}.
The perfect fluid conjecture states that the ratio of the shear viscosity $\eta$ to the entropy density $s$
has a universal minimum,
\begin{equation}
\frac{\eta}{s}\geq \frac{1}{4\pi}\frac{\hbar}{k_B}.
\label{eq:perfect}
\end{equation}
One example of a nearly perfect fluid is the quark-gluon plasma  produced in a collision between two gold ions at an energy of 100 GeV per nucleon, which is thought to be a good approximation to the state of matter that existed microseconds after the Big Bang~\cite{McClerranRHIC}. The collision produces a cigar-shaped plasma at a temperature of $2\times 10^{12}$ K that exhibits elliptic flow, where the narrow direction of the plasma expands faster than the long direction, as observed in the momentum distribution~\cite{Heinz}. A second example is a strongly interacting two-component Fermi gas of atoms, at a temperature of $10^{-7}$ K. Released from a cigar-shaped optical trap, a strongly interacting Fermi gas also exhibits elliptic flow, which is directly observed in the spatial profile of the expanding cloud~\cite{OHaraScience}. Despite a difference in temperature of 19 orders of magnitude and a difference in a density of 25 orders of magnitude, both systems exhibit nearly  frictionless hydrodynamics and have  a similar ratio $\eta/s$.

Ultracold strongly interacting Fermi gases are generally of broad interest,
as they provide a tunable tabletop paradigm  for
strongly interacting systems, ranging from high temperature
superconductors to nuclear matter. First observed in 2002, quantum degenerate, strongly
interacting Fermi gases are being widely
studied~\cite{OHaraScience,RMP2008,ZwergerReview,ZwierleinFermiReview}. To obtain strong
interactions, characterized by a divergent s-wave scattering length~\cite{p-wave}, a bias magnetic field is used to tune the gas to a broad collisional (Feshbach) resonance, where the
range of the collision potential is small compared to the interparticle spacing.
In this so-called unitary regime,  the two-body interaction potential produces no relevant length scales. Hence, the local thermodynamic and transport  properties of a resonantly interacting gas are determined by the interparticle spacing $L$ and the thermal de Broglie wavelength $\lambda_T$, i.e., they are universal functions of  the density $n$ and temperature $T$~\cite{HoUniversalThermo}. Unitary Fermi gases therefore provide a scale-invariant system  for exploring minimum viscosity hydrodynamics.

The $\eta/s$ ratio is experimentally accessible in a trapped universal Fermi gas, where the entropy and other thermodynamic properties have been measured both globally~\cite{JointScience,LuoEntropy,JinPotential,DrummondUniversal,ThermoLuo} and most recently, locally~\cite{ThermoUeda,ThermoSalomon}. As the viscosity can be determined from hydrodynamic experiments~\cite{ShuryakQuantumViscosity,BruunViscousNormalDamping,SchaferRatio,TurlapovPerfect},  the predicted minimum ratio can be directly compared to that from Fermi gas experiments~\cite{SchaferRatio,TurlapovPerfect,CaoViscosity}.

The scale of the $\eta/s$ ratio for a unitary Fermi gas can be understood using dimensional analysis. As noted above, there are only two natural length scales $l\in\{L,\lambda_T\}$. Shear viscosity has units of momentum/area. As the  natural momentum is of order $\hbar/l$ and the natural area is $l^2$, $\eta \propto \hbar/l^3$.
At temperatures well below the Fermi temperature at which
degeneracy occurs, the Fermi momentum sets the scale so
$l\simeq L$, and $\eta\propto\hbar /L^3\propto\hbar n$.
For a normal fluid above the critical temperature, the scale of entropy
density $s\simeq n\,k_B$ and  $\eta/s\simeq\hbar /k_B$.
For much higher temperatures above the Fermi temperature, one expects
that $l\simeq\lambda_T\propto\hbar T^{-1/2}$, so that the shear viscosity scales as $\eta\propto \hbar/\lambda_T^3\propto T^{3/2}/\hbar^2$.

\section{Universal Hydrodynamics}
\label{sec:universalhydro}

To properly  measure the shear viscosity with high precision over a wide temperature range, we employ two experimental methods~\cite{CaoViscosity}. To determine the shear viscosity at low temperatures, we measure the damping rate of the radial breathing mode excited by releasing the cloud for a short time and then recapturing it~\cite{KinastDampTemp}. For measurement at high temperatures, we observe the aspect ratio of the cloud as a function of time after the cloud is released from a deep optical trap. To consistently extract the viscosity from these two experiments, we use universal hydrodynamic equations, which contain both the friction force and the viscous heating rate.

We determine the shear viscosity $\eta$ in the normal fluid regime by using a hydrodynamic
description of a single component fluid, where the velocity field $\mathbf{v}(\mathbf{x},t)$ is determined by
the scalar pressure and the shear viscosity pressure tensor,
\begin{equation}
m\left(\partial_t
+\mathbf{v}\cdot\nabla\right)v_i=f_i + \sum_j\frac{\partial_j (\eta\,\sigma_{ij}+\zeta\sigma^{'}_{ij})}{n}-\partial_i U_{trap},
\label{eq:force}
\end{equation}
where $\mathbf{f}=-\nabla P/n$ is the  force per particle arising
from the scalar pressure $P$, $U_{trap}$ is the potential of the optical trap, and $m$ is the atom mass. The second term on the right describes the friction forces
arising from both shear $\eta$ and bulk $\zeta$ viscosities, where $\sigma_{ij}=\partial
v_i/\partial x_j+\partial v_j/\partial
x_i-2\delta_{ij}\nabla\cdot\mathbf{v}/3$  and $\sigma^{'}_{ij}=\delta_{ij}\nabla\cdot\mathbf{v}$.

For a unitary gas, the  evolution equation for the pressure takes
a simple form, since $P=2{\cal E}/3$~\cite{HoUniversalThermo,
ThomasUniversal}, where ${\cal E}$ is the local energy density
(sum of the kinetic and interaction energy). Then, energy conservation and Eq.~\ref{eq:force} implies~\cite{CaoViscosity}
\begin{equation}
(\partial_t +\mathbf{v}\cdot\nabla + 5\nabla\cdot\mathbf{v}/3)P=2\dot{q}/3.
\label{eq:pressure}
\end{equation}
Here, the heating rate per unit volume $\dot{q}=\eta\,\sum_{ij}\sigma_{ij}^2/2+\zeta(\nabla\cdot\mathbf{v})^2$ arises from friction forces due to relative motion (shear viscosity) and dilation (bulk viscosity) of the volume elements.

The evolution equation for the force per particle $f_i$ is readily determined by differentiating Eq.~\ref{eq:pressure} with respect to $x_i$ and using the continuity equation for the density,
\begin{equation}
 \left(\partial_t
+\mathbf{v}\cdot\nabla
+\frac{2}{3}\nabla\cdot\mathbf{v}\right)f_i + \sum_j(\partial_i v_j)f_j -\frac{5}{3}\left(\partial_i\nabla\cdot\mathbf{v}\right)\frac{P}{n}=-\frac{2}{3}\frac{\partial_i\dot{q}}{n}.
\label{eq:forceperparticle}
\end{equation}
 Force balance in the trapping potential $U_{trap}(\mathbf{x})$, just before release of the cloud, determines the initial condition $f_i(0)=\partial_i U_{trap}(\mathbf{x})$.

These hydrodynamic equations include both the force and the heating arising from shear and bulk viscosities. The solution is greatly simplified when the cloud is released from a deep, nearly harmonic trapping potential $U_{trap}$, as $f_i(0)$ is then linear in the spatial coordinate. If we neglect viscosity, the force per particle and hence the
velocity field remain linear functions of the spatial coordinates as the cloud expands.
Thus $\partial_i(\nabla\cdot\mathbf{v})=0$ and the pressure $P$
does not appear in Eq.~\ref{eq:forceperparticle}. Numerical integration~\cite{ThomasNUM} shows that non-linearities in the velocity field are very small even if the viscosity is not zero, because dissipative forces tend to restore a linear flow profile. Hence,  the evolution equations~\ref{eq:force} and~\ref{eq:forceperparticle} are only weakly dependent on the precise initial spatial profile of $P$ and  independent of the detailed thermodynamic properties.

For the expansion and breathing mode experiments, we assume that the velocity field is exactly linear in the spatial
coordinates. We take the force per particle to be of the form $f_i=a_i(t)x_i$ and assume that the density changes by a scale transformation~\cite{Menotti}, $n(\mathbf{x},t)=n[x/b_x(t),y/b_y(t),z/b_z(t)]/\Gamma$. Here $\Gamma =b_xb_yb_z$ is the volume scale factor and $\int d^3\mathbf{x}\,n(\mathbf{x},t)=N$ is the total number of atoms, which is conserved.  Current conservation then requires that the velocity field take the form $v_i=x_i\,\dot{b}_i(t)/b_i(t)$, so that the pressure term in Eq.~\ref{eq:forceperparticle} vanishes as discussed above.

The bulk viscosity is predicted to vanish in the normal fluid phase at unitarity~\cite{SonBulkViscosity,EscobedoBulkViscosity}, so we initially exclude the bulk viscosity in
our universal hydrodynamic equations~\ref{eq:force}~and~\ref{eq:forceperparticle}, to extract the shear viscosity.
In general, the (static) shear viscosity takes the universal form
\begin{equation}
\eta (\mathbf{x},t) = \alpha(\theta)\,\hbar n,
 \label{eq:eta}
 \end{equation}
 where $\theta=T/T_F(n)$ is the local reduced temperature and $T_F(n) = \hbar^2(3\pi^2 n)^{2/3}/(2mk_B)$ is the local Fermi temperature. Kinetic theory shows that $\eta\rightarrow 0$ in the low density region of the cloud~\cite{BruunViscous}, as required for energy conservation.

 With these assumptions, the evolution equation for the force per particle, Eq.~\ref{eq:forceperparticle} yields
 $$
 n\,x_i\,\left(\dot{a}_i+2\frac{\dot{b_i}}{b_i}\,a_i+\frac{2}{3}\sum_j\frac{\dot{b_j}}{b_j}\,a_i\right)=
 -\frac{2}{3}\,\partial_i\dot{q},
$$
 where, for zero bulk viscosity ($\zeta =0$), the heating rate is $\dot{q}=\eta\,\sum_{ij}\sigma_{ij}^2/2$. Here, $\sigma_{ij}$ is evaluated using $\partial_j v_i=\delta_{ij}\,\dot{b}_i/b_i$, which is spatially constant. Multiplying both sides by $x_i$ and integrating over all space, the left side yields $\int d^3\mathbf{x} \,n\, x_i^2=N\langle x_i^2\rangle=N\langle x_i^2\rangle_0\, b_i^2(t)$, where $\langle x_i^2\rangle_0$ is the equilibrium mean square size of the trapped cloud in the i$^{th}$ direction. Integrating by parts, and assuming $\eta \rightarrow 0$ as $n\rightarrow 0$, the right side is proportional to  $\int d^3\mathbf{x} \,\eta (\mathbf{x},t)$, yielding
 \begin{equation}
 \dot{a}_i+2\frac{\dot{b_i}}{b_i}\,a_i+\frac{2}{3}\sum_j\frac{\dot{b_j}}{b_j}\,a_i=\frac{1}{3}\,\frac{\int d^3\mathbf{x} \,\eta (\mathbf{x},t)}{N \langle x_i^2\rangle_0\, b_i^2(t)}\,\sum_{ij}\sigma_{ij}^2.
 \label{eq:forceperparticle1}
 \end{equation}

 Similarly, Eq.~\ref{eq:force} for the velocity field takes the form
 \begin{equation}
 \frac{\ddot{b}_i}{b_i}=\frac{a_i}{m}-\omega_i^2-\frac{\int d^3\mathbf{x} \,\eta (\mathbf{x},t)}{m\,N \langle x_i^2\rangle_0\, b_i^2(t)}\,\sigma_{ii}.
 \label{eq:force1}
 \end{equation}
 Here, the $\omega_i^2$ term on the right arises from the harmonic trapping potential, which is retained for the breathing mode and set equal to zero for expansion, where the cloud is released from the trap.

 The right hand sides of Eqs.~\ref{eq:forceperparticle1}~and~\ref{eq:force1} depend on the trap-averaged viscosity parameter, $\bar{\alpha}$, where
  \begin{equation}
  \bar{\alpha}\equiv\frac{1}{N\hbar}\int d^3\mathbf{x}\,\eta(\mathbf{x},t).
 \label{eq:viscositycoeff}
 \end{equation}
 The spatial integral exists, since as discussed above, as $\eta\rightarrow 0$ as the density goes to zero at the edges of the cloud~\cite{BruunViscous}.  As the viscosity produces a first order perturbation to perfect hydrodynamics, we can evaluate $\theta$ in Eq.~\ref{eq:eta} using a zeroth order adiabatic approximation, so that $\theta$ has a zero convective derivative everywhere. Since the number of atoms in a volume element is conserved along a stream tube, $\bar{\alpha}$ is then time-independent.

 For measurements of the viscosity at high temperatures, the cloud is released from a deep optical trap and the aspect ratio $\sigma_x(t)/\sigma_z(t)=\sqrt{\langle x^2\rangle/\langle z^2\rangle}$ is observed as a function of time after release, Fig.~\ref{fig:aspectratio}. Here, $x$ and $z$ refer to the initially narrow and long directions of the trapped cloud, respectively. Eqs.~\ref{eq:forceperparticle1}~and~\ref{eq:force1} are solved using the measured trap frequencies and the initial mean square cloud sizes, with the initial conditions $b_i(0)=1$, $\dot{b}_i(0)=0$, and $a_i(0)=m\omega_i^2$. The ratio $\sigma_x(t)/\sigma_z(t)=(\omega_z/\omega_x)\,b_x(t)/b_z(t)$ is determined as a function of time and compared to the data, yielding very good fits, Fig.~\ref{fig:aspectratio}, with $\bar{\alpha}$  as the only free parameter, which is determined by minimizing $\chi^2$.

  We measure the viscosity at low temperature using a breathing mode~\cite{KinastDampTemp}, which is excited by  a brief release and subsequent recapture. The  oscillation of the transverse radius of the trapped cloud is observed as a function of time after excitation and the damping rate is measured by fitting an exponentially damped sinusoid to the data.
 For the breathing mode, the amplitude is small and the cloud radii change by a scale transformation of the form $b_i=1+\epsilon_i$, with $\epsilon_i<<1$. As the heating term in Eq.~\ref{eq:forceperparticle1} containing $\sigma_{ij}^2$ is  $\propto\dot{\epsilon}_i^2$, the heating rate is negligible for the breathing mode and the force per particle evolves adiabatically to first order in small quantities, so that  Eq.~\ref{eq:forceperparticle1} yields $a_i=m\omega_i^2/(b_i^2\Gamma^{2/3})$. Using this in Eq.~\ref{eq:force1}, we determine the breathing frequency  $\omega_B$ and the damping rate $1/\tau$. For a cylindrically symmetric trap, where $\omega_x=\omega_y=\omega_\perp$, $\omega_B=\omega_\perp\, \sqrt{10/3}$. The damping rate  arises  from the  $\sigma_{ii}$ term, which, for a cylindrically symmetric trap is proportional to  $\dot{\epsilon}_x=\dot{\epsilon}_y =\dot{\epsilon}_\perp$.

 More generally, we assume a nearly cylindrical trap potential as used in the experiments, with $\omega_x,\omega_y$ the transverse trap frequencies and $\delta\equiv(\omega_x-\omega_y)/\sqrt{\omega_x\omega_y}<<1$. Then, $\omega_B=\sqrt{10\omega_x\omega_y/3}$. In terms of the cloud size observed for the x-direction, the damping rate is given by
 \begin{equation}
 \frac{1}{\tau}=\frac{\hbar\bar{\alpha}}{3m\langle x^2\rangle_0}\left(1-\delta\right).
 \label{eq:dampingrate}
 \end{equation}
    The transverse mean square size of the trapped cloud before excitation is given by $\langle x^2\rangle_0=\langle x^2\rangle/b_x^2(t)$, where $\langle x^2\rangle$ is determined by imaging at a time $t$ after the cloud is released and $b_x(t)$ is the calculated hydrodynamic expansion factor.  We self-consistently determine $b_x(t)$ and $\bar{\alpha}$ from the measured damping rate and the measured cloud size after expansion, using  Eq.~\ref{eq:dampingrate} as a constraint. Since the viscosity is small at low temperatures, we initially calculate $b_x(t)$ for zero viscosity to obtain an initial guess for $\langle x^2\rangle_0$. This yields an initial approximation for $\bar{\alpha}$. This initial value is then used in Eqs.~\ref{eq:forceperparticle1}~and~\ref{eq:force1} (with $\omega_i^2=0$ and including the heating rate) to determine a better approximation to $b_x(t)$, which in turn yields a better value for $\langle x^2\rangle_0$ and $\bar{\alpha}$. This procedure quickly converges.

\section{Experiments}

Our experiments employ a 50-50 mixture of the two lowest hyperfine
states of $^6$Li fermions, which is confined in a stable CO$_2$
laser trap~\cite{OHaraStable}.  The mixture exhibits a Feshbach
resonance at 834 G for which the s-wave scattering length
diverges~\cite{BartensteinFeshbach}, producing a unitary gas where
the two-body scattering cross section is inversely proportional to the
relative kinetic energy.

After forced evaporation by lowering the trap depth, the trap is recompressed.
At the final trap depth, parametric resonance is used to measure the
oscillation frequencies of weakly interacting atoms. The
frequencies obtained from the measurements are corrected for
anharmonicity to determine the harmonic oscillation frequencies
for energies small compared to the trap depth. For the deep optical trap used for the high temperature measurements, we obtain $\omega_z
= 2\pi\times (182.7\pm 0.5$) Hz, $\omega_x = 2\pi\times (5283\pm
10)$ Hz, $\omega_y = 2\pi\times (5052\pm 10)$ Hz, and
$\bar{\omega}=(\omega_x\omega_y\omega_z)^{1/3}=2\pi\times (1696\pm
9)$  Hz. The total number of atoms ranges from $N = 4.0\times 10^5$ at $E=2.3\,E_F$ to $N=6.0\times 10^5$ at $E=4.6\,E_F$,
where less evaporation is employed.
For $N = 6.0\times 10^5$, the Fermi energy of an ideal gas at the trap
center is $E_F=(3N)^{1/3}\,\hbar\bar{\omega}= k_B \times
9.9\,\mu$K, small compared to the trap depth $U_0 = k_B\times
460\,\mu$K.

\begin{figure}[h]
\begin{center}\
(A)
\includegraphics[width=150mm]{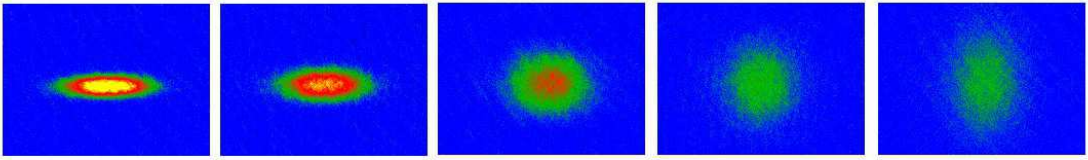}

(B)
\includegraphics[width=120mm]{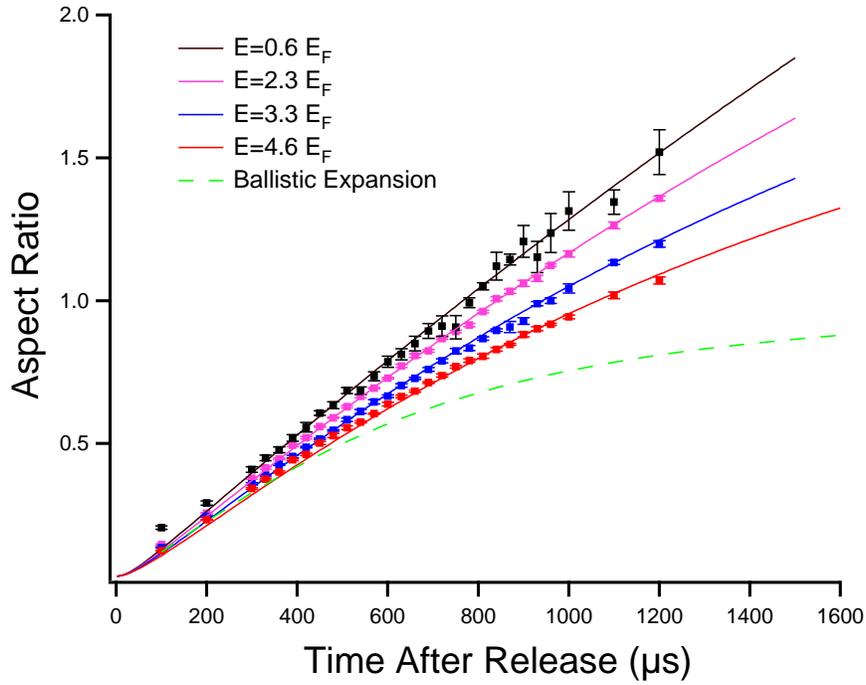}
\end{center}
\caption{Anisotropic expansion. (A) Cloud absorption images for 0.2, 0  .3, 0.6, 0.9, 1.2 ms expansion time, $E=2.3\,E_F$; (B) Aspect ratio versus time. The expansion rate decreases at higher energy as the viscosity increases. Solid curves: Hydrodynamic theory with the viscosity as the fit parameter.  Error bars denote statistical fluctuations in the aspect ratio.\label{fig:aspectratio}}
\end{figure}

For the high temperature experiments, the  aspect ratio $\sigma_x(t)/\sigma_z(t)$ is measured as a
function of time after release to determine $\bar{\alpha}$ for  energies $E$ between $2.3\,E_F$
and $4.6\,E_F$, Fig.~\ref{fig:aspectratio}. We also take expansion data at one low energy point $E=0.6\, E_F$, where the viscosity is small compared to that obtained at higher temperatures and the density profile is approximately a zero temperature Thomas-Fermi distribution. The black curve shows the fit for zero viscosity and no free parameters. To obtain a high signal to background ratio, we measure the aspect ratio only up to 1.4. For comparison, the green dashed curve shows the prediction for a ballistic gas.
In these experiments, where the total energy of the gas $E$ larger than $2\,E_F$,  we find that the density profile is well fit by a Gaussian $n(x,y,z,t)=n_0(t)\,\exp(-x^2/\sigma_x^2-
y^2/\sigma_y^2-z^2/\sigma_z^2)$, where $\sigma_i(t)=b_i(t)\sigma_i(0)$ is a time
dependent width, $n_0(t)=N/(\pi^{3/2}\sigma_x\sigma_y\sigma_z)$ is the central
density, and $N$ is the total number of atoms.

For the breathing mode experiments, the gas is initially cooled to nearly the ground state, where  the total number of atoms is $N=2.0\times 10^5$, and then the trap is recompressed. Energy is then added by release and recapture, the gas is allowed to equilibrate, and the breathing mode is excited. The damping rate is measured for energies $E$ between $0.5\,E_F$ and $2.5\,E_F$. For the shallow trap used in the breathing mode experiments, the measured parametric resonance frequencies are: $\omega_\perp=\sqrt{\omega_x\omega_y} = 2\pi\times
1696(10)$ Hz, $\omega_x/\omega_y=1.107(0.004)$, and
$\omega_z=2\pi\times 71(3)$ Hz, so that
$\bar{\omega}=(\omega_x\omega_y\omega_z)^{1/3}=2\pi\times 589(5)$
Hz is the mean oscillation frequency and $\lambda
=\omega_z/\omega_\perp=0.045$ is the anisotropy parameter. The
typical Fermi temperature $T_F=(3 N)^{1/3}\hbar\bar{\omega}/k_B$ of a
corresponding noninteracting gas is $\simeq 2.4\,\mu$K, and
the trap depth is $U_0/k_B=35\,\mu$K. The transverse mean square size $\langle x^2\rangle$ is measured after an expansion time $t=1$ ms.

The initial energy per particle $E$  is measured from the density profile of the trapped cloud along any one direction by exploiting the virial theorem~\cite{ThermoLuo,ThomasUniversal}, which holds in the
unitary regime. For measurements of the axial density profile,
\begin{equation}
E=3m\omega_z^2\langle z^2\rangle[1-\kappa\,\langle z^2\rangle/\sigma_{Fz}^2],
\label{eq:energy}
\end{equation}
where $\kappa=15E_F/(4U_0)$, corrects for anharmonicity in the
trapping potential~\cite{ThermoLuo}.
Here $\langle z^2\rangle$ is the mean square size of the axial density profile of the
trapped cloud and $\sigma_{Fz}=\sqrt{2E_F/(m\omega_z^2)}$ is the Fermi radius for the z-direction. The energy also can be measured from the mean square size in the transverse direction, i.e., $z\rightarrow x$. A systematic uncertainty of 3\% in $E_F$ arises from the $\leq 10$\% uncertainty in the absolute atom number $N$~\cite{JosephSound}.

 As shown in Fig.~\ref{fig:aspectratio}, the expansion data are very well fit over the range of energies studied, using $\bar{\alpha}$ as the only free parameter.  We find that the friction force produces a curvature that matches the aspect ratio versus time data. For the expansion data, the indirect effect of heating is significant in increasing the outward force. Including heating in the analysis significantly increases the fitted $\bar{\alpha}$ compared to that obtained when heating is omitted~\cite{CaoViscosity}, as discussed below. For the breathing mode experiments at low temperature, the effect of heating is small, as discussed above.

\begin{figure}
\begin{center}\
\includegraphics[width=120mm]{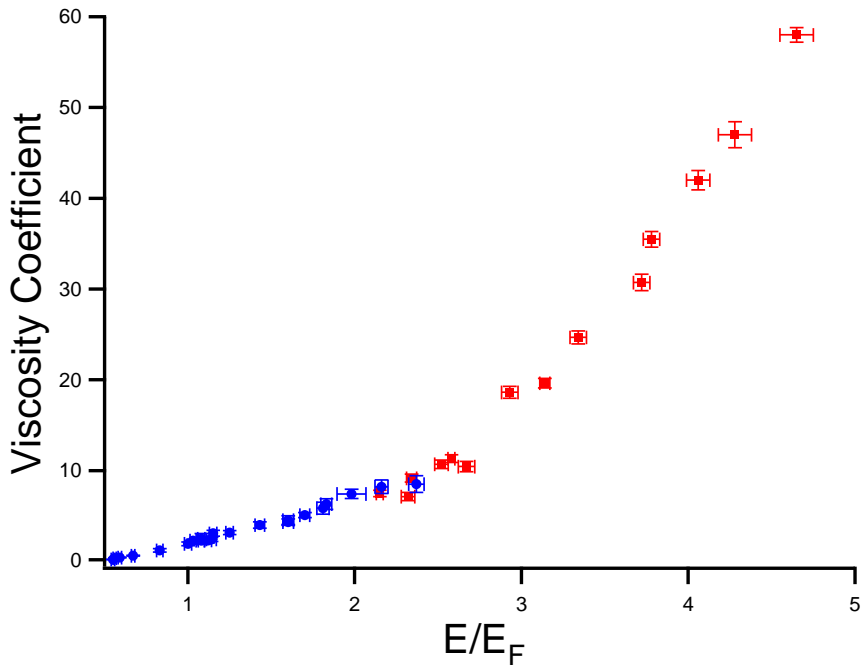}
\end{center}
\caption{Trap-averaged viscosity coefficient $\bar{\alpha}=\int d^3\mathbf{x}\,\eta/(\hbar N)$ versus initial energy per atom. Blue circles: Breathing mode measurements; Red squares: Anisotropic expansion measurements. Bars denote statistical error arising from the uncertainty in $E$ and the cloud dimensions.  \label{fig:viscosity}}
\end{figure}

\begin{figure}
\begin{center}\
\includegraphics[width=120mm]{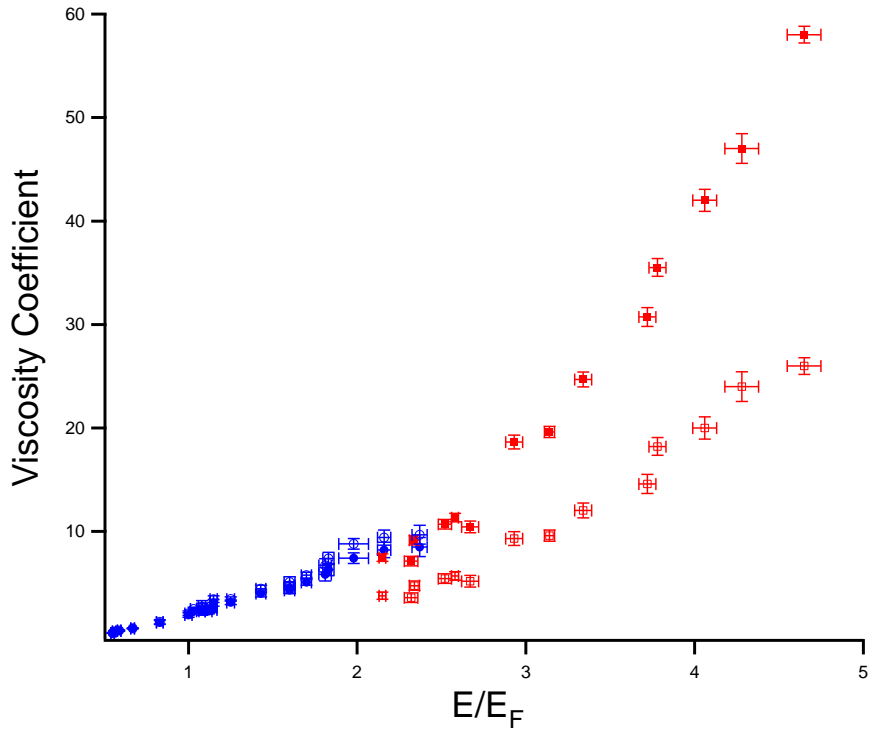}
\end{center}
\caption{Effect of the heating rate in Eq.~\ref{eq:forceperparticle} on the measured viscosity coefficient $\bar{\alpha}$ versus initial energy per atom. Solid(open) circles/squares: heating is included(excluded). Blue circles: low energy breathing mode data; Red squares: high energy expansion data.  The high and low temperature data smoothly join only when heating is included. Bars denote errors arising from the uncertainty in $E$ and the cloud dimensions.\label{fig:heating}}
\end{figure}

Together, the breathing mode and expansion measurements determine the fitted viscosity coefficients $\bar{\alpha}$ for the entire energy range, as shown in Fig.~\ref{fig:viscosity}.  As shown in \S~\ref{sec:ReducedTemp}, by calibrating the temperature, we determine $\bar{\alpha}$ as function of reduced temperature $\theta_0$ at the trap center, prior to release. These results for the trap-averaged viscosity coefficient as a function of reduced temperature can be used to test predictions~\cite{LevinViscosity,TaylorViscosity,ZwergerViscosity}, within the local density approximation.

Despite the large values of $\bar{\alpha}$ at the higher energies, the viscosity causes only a moderate perturbation to the adiabatic expansion, as shown by the expansion data and the fits in Fig.~\ref{fig:aspectratio}. The breathing mode data and expansion data smoothly join, provided that the heating rate is included in the analysis. In contrast, omitting the heating rate  produces a significant discontinuity, nearly a factor  of 2,  between the high and low temperature viscosity data, as shown in Fig.~\ref{fig:heating}. The agreement between these very different measurements when heating is included shows that hydrodynamics in the universal regime is well described  by Eqs.~\ref{eq:force} and~\ref{eq:forceperparticle}.

\section{Vanishing of the Bulk Viscosity at Unitarity}
 The bulk viscosity is predicted to vanish in the normal fluid phase at unitarity~\cite{SonBulkViscosity,EscobedoBulkViscosity}, which is why we did not include the bulk viscosity in
 our initial analysis to extract the shear viscosity. Energy conservation also requires the bulk viscosity to vanish in the two-body collision limit. We show that by including the bulk viscosity in both heating and force equations, the best fit to our anisotropic expansion data is the one for which the bulk viscosity exactly vanishes.
\begin{figure}
\begin{center}\
\includegraphics[width=120mm]{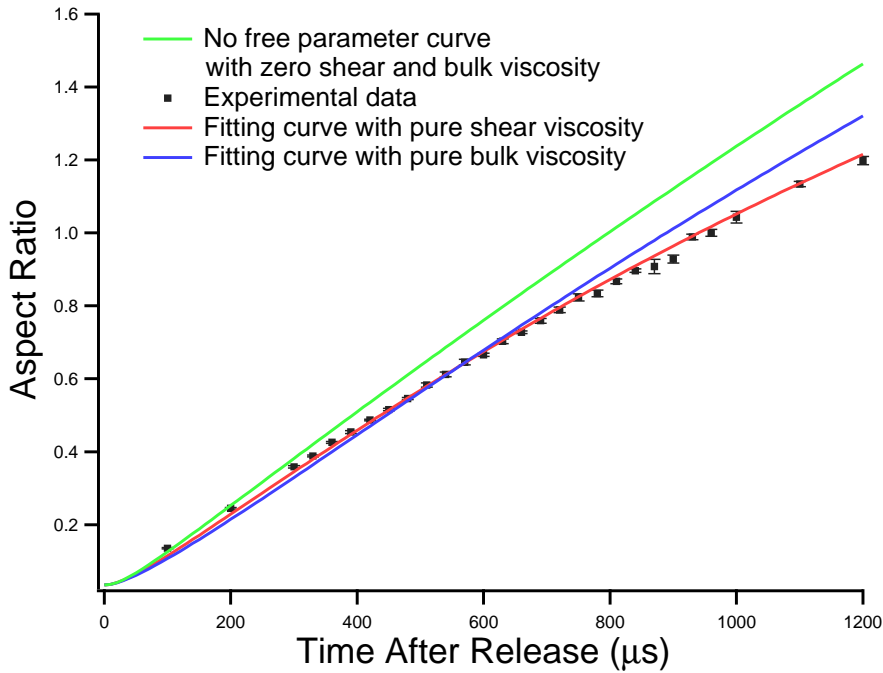}
\end{center}
\caption{Comparison between the best $\chi^2$ fit with pure shear viscosity and the best $\chi^2$ fit with pure bulk viscosity. The green curve is the no free parameter prediction with zero shear and bulk viscosity. The black dots are the anisotropic expansion data at $E/E_{F}=3.3$. Error bars denote statistical fluctuations in the aspect ratio. The red curve is the best $\chi^2$ fit with pure shear viscosity coefficient $\bar\alpha_{S}=24.4$ and reduced $\chi^2=1.6$. The blue curve is the best $\chi^2$ fit with pure bulk viscosity coefficient $\bar\alpha_{B}=16.7$ and reduced $\chi^2=8.6$.\label{fig:Bulkshearcomparison}}
\end{figure}

\begin{figure}
\begin{center}\
\includegraphics[width=120mm]{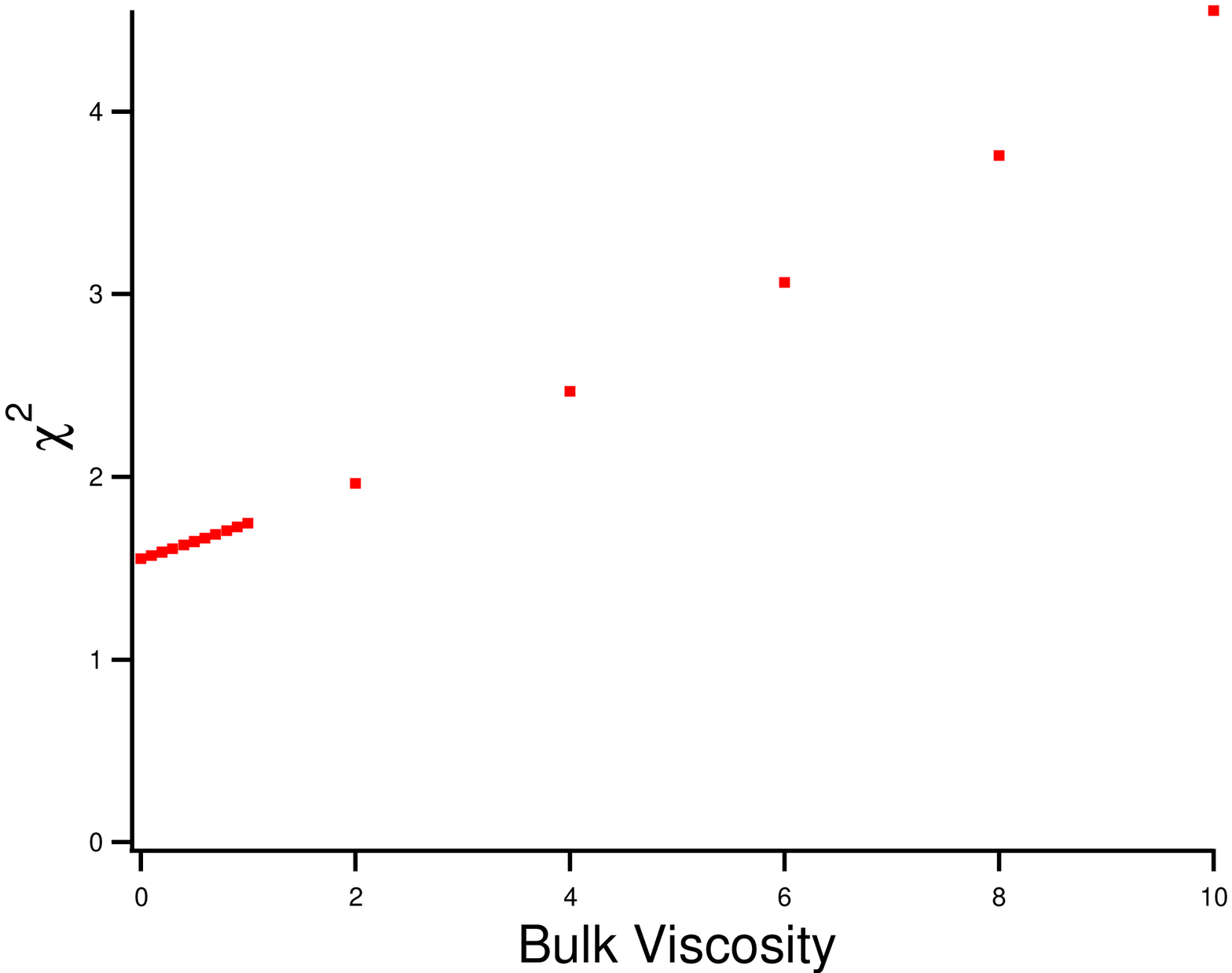}
\end{center}
\caption{$\chi^2$ per degree of freedom versus bulk viscosity with shear viscosity as the only free parameter.\label{fig:BulkVersusChisquare}}
\end{figure}

By including the bulk viscosity in Eqs.~\ref{eq:force} and~\ref{eq:forceperparticle}, with the same forms as used above for the density profile, the velocity field and the force per particle, we can apply a $\chi^2$ fit to our anisotropic expansion data. As shown in Fig.~\ref{fig:Bulkshearcomparison}, not only is the reduced $\chi^2$ of the pure bulk viscosity fit much larger than that of the pure shear viscosity fit, but also it does not produce the curvature that our experimental data  shows. This curvature arises from the fact that the anisotropic shear viscosity pressure tensor slows the transverse expansion and speeds up the axial expansion. In contrast, the bulk viscosity pressure tensor symmetrically slows the expansion in all directions.

We also choose the bulk viscosity and apply a $\chi^2$ fit to our data using the shear viscosity as the only free parameter~\cite{Shaeferbulksuggestion}. As shown in Fig.~\ref{fig:BulkVersusChisquare}, zero bulk viscosity gives the minimum reduced $\chi^2$, which is consistent with the prediction of vanishing bulk viscosity in the normal fluid phase at unitarity~\cite{SonBulkViscosity,EscobedoBulkViscosity}.

\section{Shear Viscosity versus Reduced Temperature}
\label{sec:ReducedTemp}

For comparison with predictions for the temperature dependence of the shear viscosity~\cite{BruunViscousNormalDamping,LevinViscosity,TaylorViscosity,ZwergerViscosity}, we give the trap-averaged viscosity coefficient $\bar{\alpha}$ as a function of reduced temperature $\theta_0$ at the trap {\it center}, prior to release of the cloud,
\begin{equation}
\theta_0 = \frac{T_0}{T_F(n_0)} = \frac{T_0}{T_{FI}}\left(\frac{n_I}{n_0}\right)^{2/3}.
\label{eq:theta}
\end{equation}
Here, the local Fermi temperature at the trap center, before the cloud is released, is given by $T_F(n_0) = \hbar^2
(3\pi^2n_0)^{2/3}/(2mk_B)$ and $T_{FI} = E_F /k_B = T_F (n_I)$ is
the ideal gas Fermi temperature at the trap center, with  $n_I$
the ideal gas central density for a zero temperature Thomas-Fermi
distribution, $8N/(\pi^2\sigma_{Fx}\sigma_{Fy}\sigma_{Fz})$. As shown previously, $(n_I/n_0)^{2/3}\propto E/E_F$. Hence, at higher temperatures, where  $T_0/T_F\propto E/E_F$, we have $\theta_0\propto (E/E_F)^2$~\cite{CaoViscosity}.

To measure $T_0/T_{FI}$ for the  high temperature expansion data, where $E>2E_F$, we use the second virial coefficient approximation to the local energy density~\cite{HoMuellerHighTemp},
\begin{equation}
{\cal E}=(3/2)nk_BT(1+B_2\,n\lambda_T^3),
 \label{eq:energydensity}
 \end{equation}
 where $n$ is the total density and $\lambda_T=h/\sqrt{2\pi m k_BT}$ is the thermal wavelength. $B_2=1/2^{7/2}-b_2/\sqrt{2}$ is the second virial coefficient for a unitary gas, with $b_2=1/2$, which is universal and known to be accurate for experiments in
this temperature regime~\cite{DrummondUniversal,ThermoSalomon}.
Here, the first term in $B_2$ arises from degeneracy for each spin state of the 50-50 mixture and the second term
arises from interactions between them. Force balance in the
trap requires~\cite{ThomasUniversal,ThermoLuo} $\int d^3\mathbf{x}\,{\cal E}=\int
d^3\mathbf{x}\,\mathbf{x}\cdot\nabla U_{trap}(\mathbf{x})/2$.
For the Gaussian density profile observed in the high temperature experiments, one obtains
\begin{equation}
\frac{T_0}{T_{FI}}=\frac{\sigma_z^2}{\sigma_{Fz}^2}\left[1-\frac{5}{2}\frac{E_F}{U_0}\frac{\sigma_z^2}{\sigma_{Fz}^2}
-\frac{B_2}{6\sqrt{2}}\,
\frac{\sigma_{Fz}^6}{\sigma_{z}^6}\right]. \label{eq:temp}
\end{equation}
 Note that Eq.~\ref{eq:temp} includes first order corrections arising from $B_2$ and trap anharmonicity.
 For the high temperature data, where the density profile is Gaussian, we have $(n_I/n_0)^{2/3} = 4(\sigma^2_z/\sigma^2_{Fz})/
\pi^{1/3}$. Using the initial $T_0/T_{FI}$ obtained from the cloud profile, we determine $\theta_0$ from Eq.~\ref{eq:theta}.
 \begin{figure}
\begin{center}\
\includegraphics[width=120mm]{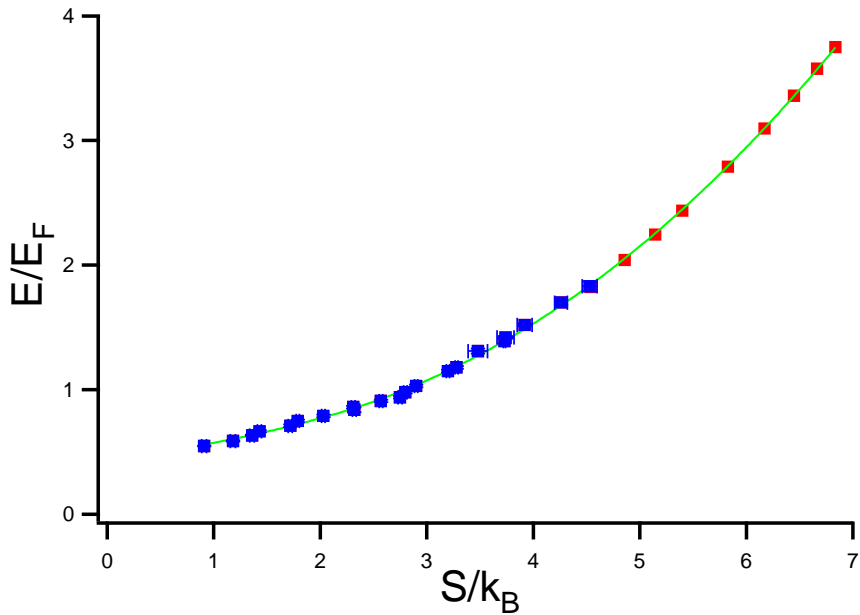}
\end{center}
\caption{Total energy per particle of a strongly interacting Fermi
gas at $840$ G versus the entropy per particle. The blue dots show
the entropy obtained by adiabatically sweeping the magnetic field from $840G$ to $1200G$  and using an exact many-body calculation~\cite{DrummondUniversal} for the entropy at 1200 G, where $k_Fa=-0.75$. [See $S^*_{1200}$ in Table 1, Ref.~\cite{ThermoLuo}].  The red dots are the theoretical calculations using the
second virial coefficient approximation including the trap
anharmonicity for the trap condition of Ref.~\cite{LuoEntropy}. The green curve is the power law fit with a
discontinuous heat capacity, as described in Eq.~\ref{eq:EvsSlowengtwopower1}. \label{fig:energyvsentropy}}
\end{figure}

\begin{figure}
\begin{center}\
\includegraphics[width=120mm]{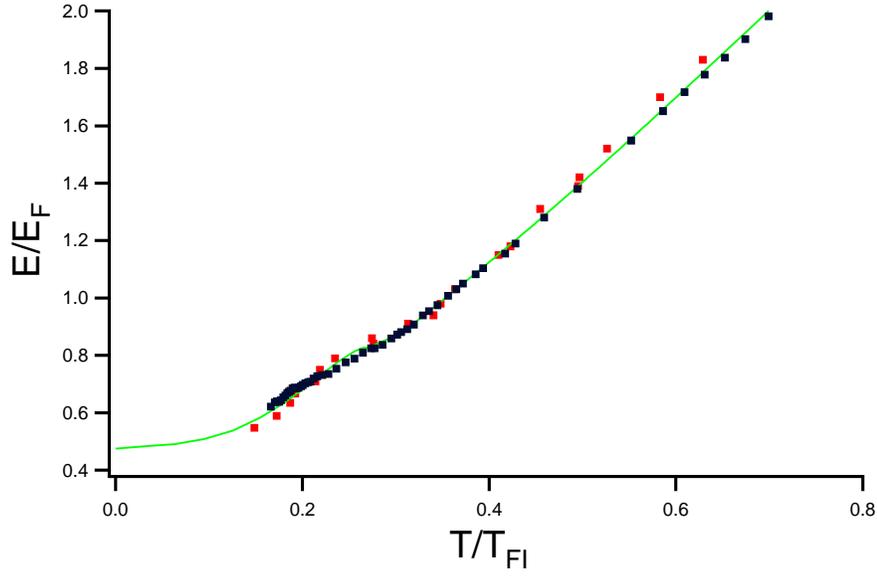}
\end{center}
\caption{Measured energy versus the temperature obtained from the calibration of Eq.~\ref{eq:EvsSlowengtwopower1} (red dots); For comparison, we show the
data obtained by the ENS group~\cite{ThermoSalomon} (black dots) and the theory of Hu et al.~\cite{HuNJPUniversal} (green curve). \label{fig:energyvsT}}
\end{figure}

  We find $T_0/T_{FI}$ for the low temperature breathing mode data from our previous measurements of energy versus entropy, for which the trap is as shallow as that used for the low temperature breathing mode measurements. The low temperature $E$ versus $S$ data employs the entropy $S^*_{1200}$ given in Table 1 of Ref.~\cite{ThermoLuo}, which is corrected for the finite interaction strength at 1200 G~\cite{IdealS}. To properly calibrate the temperature with high precision over a wide range, we join the the experimental $E(S)$ data   with theoretical calculations of $E$ and $S$ in the high temperature regime by exploiting the second virial coefficient approximation for a unitary Fermi gas, as shown in Fig.~\ref{fig:energyvsentropy}. For simplicity, we fit $E(S)$ data using a smooth curve with a discontinuous heat capacity~\cite{ThermoLuo} as follows:
\begin{eqnarray}
E_<(S)&=&E_0+\,aS^b;\,\,\,\,0\leq S\leq S_c\nonumber\\
E_>(S)&=&E_1+\,cS^d;\,\,\,\,S\geq S_c.\label{eq:EvsSlowengtwopower0}
\end{eqnarray}
Constraining the values of $E_1$ and $c$ by demanding that energy and temperature be continuous at the joining point $S_c$, one obtains:
\begin{eqnarray}
E_<(S)&=&E_0+\,aS^b;\,\,\,\,0\leq S\leq S_c\nonumber\\
E_>(S)&=&E_0+\,aS_c^b[1-b/d+b/d(S/S_c)^d];\,\,\,\,S\geq S_c.\label{eq:EvsSlowengtwopower1}
\end{eqnarray}
We fix the ground state energy $E_0=0.47$ by using $\beta=-0.60$~\cite{ThermoLuo}. Then we use a $\chi^2$ fit, where we give equal weighting to the low temperature data points and the calculated high temperature points.  We find $a=0.10(1)$, $b=1.57(15)$, $d=2.23(3)$. The critical parameters obtained from the fit are  $S_c=2.04(39)$,  $E_c=0.78(12)\,E_F$. Then, $T=\partial E/\partial S$ yields $T_c=0.24(8)\,T_{FI}$. Note that the relatively large error bar for $S_c$ and hence $T_c$ arises from joining measured data with the calculated high temperature points, and fitting $E_>(S)$ by a single power law. However, this method yields a smooth temperature calibration that reproduces the temperatures used in the virial calculation to better than 3\%, Fig.~\ref{fig:energyvsentropy}.
Then, we obtain $E/E_F$ as a function of $T_0/T_{FI}$, which yields $T_0/T_{FI}$ from the measured initial energy of the cloud, Fig.~\ref{fig:energyvsT}. The central density $n_0$, prior to release is determined from column density of the trapped cloud by fitting the spatial profile with a Gaussian distribution, which is adequate except at the lowest temperature where the density profile is a zero-temperature Thomas-Fermi Distribution.

\begin{figure}
\begin{center}\
\includegraphics[width=120mm]{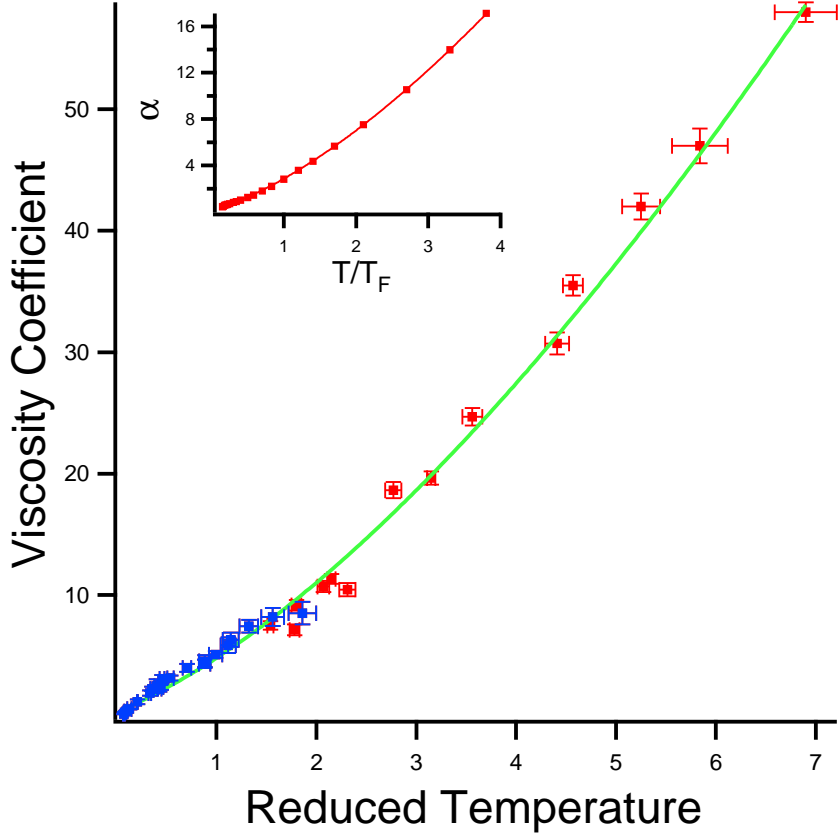}
\end{center}
\caption{Trap-averaged viscosity coefficient $\bar{\alpha}=\int
d^3\mathbf{x}\,\eta/(\hbar N)$ versus reduced temperature
$\theta_0=T_0/T_F(n_0)$ at the trap center, prior to release. Blue
dots: Breathing mode measurements; Red dots: Anisotropic expansion
measurements. Bars denote statistical errors arising from the
uncertainty in $\bar\alpha$, $E$ and the cloud dimensions. The green curve shows
the fit $\bar\alpha=\bar\alpha_{3/2}\,\theta_0^{3/2}+\bar\alpha_{1/2}\,\theta_0^{1/2}$ with
$\bar\alpha_{3/2}=2.96(3)$ and $\bar\alpha_{1/2}=1.87(8)$, for the temperature from nearly the superfluid transition point up to the two-body unitary regime. The inset shows the theoretical prediction for the local shear viscosity $\alpha$ in the normal fluid phase versus the local reduced temperature $\theta=T/T_F(n)$ from Ref.~\cite{ZwergerViscosity}. The red curve shows the fit $\alpha=\alpha_{3/2}\,\theta^{3/2}+\alpha_{1/2}\,\theta^{1/2}$ with $\alpha_{3/2}=2.11$ and $\alpha_{1/2}=0.74$.\label{fig:viscosityvstheta}}
\end{figure}

Fig.~\ref{fig:viscosityvstheta} shows the trap-averaged viscosity
coefficient $\bar\alpha$ versus the initial reduced temperature at
the trap center $\theta_0$, from nearly the ground state to the unitary two-body regime. The inset in
Fig.~\ref{fig:viscosityvstheta} is the predicted local shear viscosity in the normal fluid regime versus the local reduced temperature~\cite{ZwergerViscosity}.

\subsection{Temperature Scaling}
In the universal regime, the local viscosity at high temperatures is expected to scale as $T^{3/2}$, as described in \S~\ref{sec:intro}. The viscosity at the trap center prior to release of the cloud can be written as $\eta_0=\alpha_0\hbar\,n_0$. In the high temperature limit, we then have
\begin{equation}
\alpha_0 = \alpha_{3/2}\, \theta_0^{3/2},
\label{eq:scaling}
\end{equation}
where $\alpha_{3/2}$ is a universal coefficient~\cite{BruunViscousNormalDamping}.  As  $\theta$ has a zero convective
derivative everywhere (in the zeroth order adiabatic
approximation), $\theta_0$ at the trap center where $\mathbf{v}=0$, has a zero time
derivative and $\alpha_0$ is therefore constant as is the trap-averaged viscosity coefficient $\bar{\alpha}$.

We note that in the hydrodynamic regime, the local value of $\eta\propto T^{3/2}$ in the high temperature limit is independent of density and spatially constant, so that $\bar{\alpha}$ formally does not exist.
To test the prediction of the $T^{3/2}$ temperature scaling in the
high temperature regime, we assume that $\eta$ relaxes to the
equilibrium value in the center of the trap, but vanishes in the low
density region so that $\bar\alpha$ is well defined. As noted in \S~\ref{sec:universalhydro}, this behavior
is predicted by kinetic theory~\cite{BruunViscous} and is required for energy conservation. Assuming $\bar{\alpha}\simeq\alpha_0$, we observe the predicted $T^{3/2}$ temperature scaling in the high temperature regime~\cite{CaoViscosity}. In this case,  fitting the data with $\bar{\alpha} = \bar{\alpha}_{3/2}\, \theta_0^{3/2}$ yields  $\bar{\alpha}_{3/2}=3.4$. This is consistent with predictions, where kinetic theory~\cite{Schaefer:2009px} shows that the trap-averaged $T^{3/2}$ coefficient should be larger by a factor of $\simeq 1.3$ than that of the local value for the two-body unitary regime, $\alpha_{3/2}=2.77$~\cite{BruunViscousNormalDamping}.
However, more work is needed to fully understand the relation between the local viscosity and the trap-averaged viscosity parameter measured in the experiments.

In Fig.~\ref{fig:viscosityvstheta}, both the experimental data and the theoretical calculations are found to scale as $T^{3/2}$  in the high temperature regime, consistent with Eq.~\ref{eq:scaling}. However, we observe that the damping rate of the radial breathing mode  reaches a plateau at the higher temperatures~\cite{TurlapovPerfect,KinastDampTemp}. This flattening can be explained if $\bar{\alpha}$ has a term $\propto E$, since $1/\tau\propto\bar{\alpha}/E$, according to Eq.~\ref{eq:dampingrate}. Since $\theta_0^{1/2}\propto E$ at higher temperatures, we fit the data of Fig.~\ref{fig:viscosityvstheta}  with the two-parameter fit function,
\begin{equation}
{\bar\alpha} = \bar\alpha_{3/2}\, \theta_{0}^{3/2}+ \bar\alpha_{1/2}\, \theta_{0}^{1/2},
\label{eq:scalingforallregiondata}
\end{equation}
where $\bar\alpha$ is the trap averaged viscosity and $\theta_0$ is the local reduced temperature at the trap center.
Fitting all of the data for the normal fluid regime ($E>0.7\,E_F$), we find $\bar\alpha_{3/2}=2.96(3)$ and $\bar\alpha_{1/2}=1.87(8)$, where the errors are statistical from the fit. In this case,  $\bar{\alpha}_{3/2}=2.96$ is closer to the local value predicted for the high temperature two-body unitary regime, where $\alpha_{3/2}=2.77$.

We can also use the same two-parameter fit for the predicted local viscosity in the normal fluid regime~\cite{ZwergerViscosity}, inset Fig.~\ref{fig:viscosityvstheta},
\begin{equation}
{\alpha} = \alpha_{3/2}\, \theta^{3/2}+ \alpha_{1/2}\, \theta^{1/2},
\label{eq:scalingforallregion}
\end{equation}
where $\alpha$ is the local shear viscosity and $\theta=T/T_F(n)$ is the local reduced temperature.
We obtain $\alpha_{3/2}=2.11$ and $\alpha_{1/2}=0.74$.

The coefficient  $\bar\alpha_{1/2}=1.87$ obtained from the data is much larger than the local value obtained from the prediction
$\alpha_{1/2}=0.74$. However, to compare these parameters, we need the trap-average of the reduced temperature, $\langle \theta^{1/2}\rangle=1.84\,\theta_0^{1/2}$.  Hence, the fit to the calculated viscosity yields $\bar\alpha_{1/2}=
1.84\,\alpha_{1/2}=1.84\times0.74=1.36$, somewhat lower than the fit to the data. However, if we scale the fit coefficients for the predicted viscosity by a factor $2.77/2.11=1.31$, then $\alpha_{3/2}$  agrees by construction with the accepted high temperature two-body limit,  while the predicted $\bar\alpha_{1/2}$ increases to $1.31\times 1.36=1.78$, in good agreement with the measurements.

In the low temperature normal fluid regime, Fermi liquid theory~\cite{BruunFermiLiquid} predicts that the local viscosity  should have an upturn $\propto 1/\theta^2$. After trap-averaging, $\langle \theta^{-2}\rangle$, we find the predicted  viscosity near the critical temperature is about a factor of 2 larger than the observed value, with a temperature dependence that is not consistent with our data. In earlier work~\cite{BruunViscousNormalDamping}, pair formation  was suggested as a means of avoiding Pauli blocking, which would otherwise cause the viscosity to increase with decreasing temperature. Recent predictions~\cite{LevinViscosity}  suggest that the onset of a normal state pairing gap and the formation of pairs may suppress the viscosity at low temperature, which is consistent with our observations.

\subsection{Ratio of the shear viscosity to the entropy density}
To estimate the ratio $\eta/s$, we use the approximation
$\eta/s=\alpha\hbar n/s=(\hbar/k_B)\alpha/(s/nk_B)\simeq(\hbar/k_B)\bar{\alpha}/S$, where $S$ is the average entropy per particle of the trapped gas in units of $k_B$. For the low temperature data, we obtain $S$ from the fitting curve in Fig.~\ref{fig:energyvsentropy}.

\begin{figure}
\begin{center}\
\includegraphics[width=120mm]{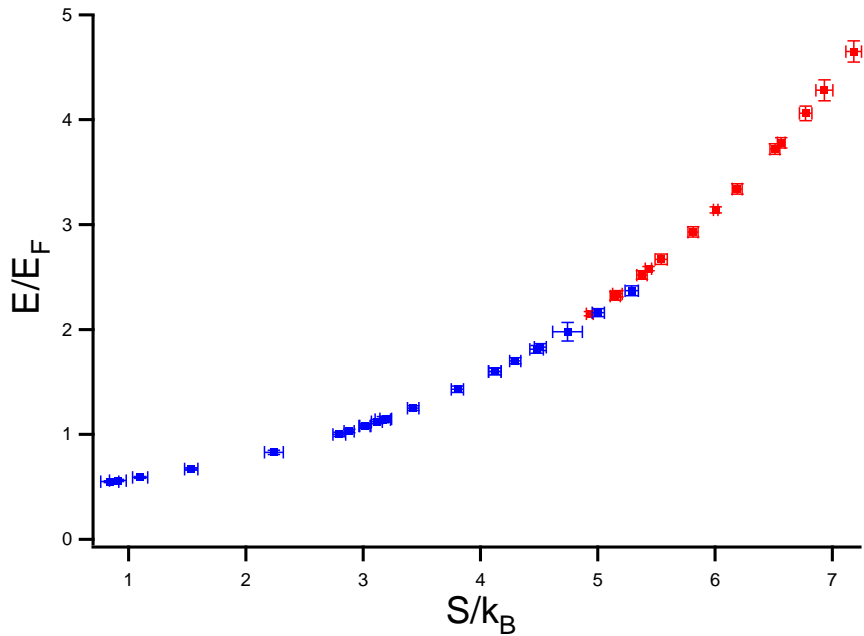}
\end{center}
\caption{Calculated trap-averaged entropy per particle versus average energy per particle for the trapped gas. Blue: Low temperature data calculated from the fit in Fig.~\ref{fig:energyvsentropy}, where error bars arise from the energy uncertainty. Red: High temperature calculation for the deep trap used in the expansion experiments, based on the second virial coefficient. Error bars arise from the energy uncertainty.\label{fig:entropy}}
\end{figure}

To determine $S$ for the high temperature data, we employ the second virial coefficient approximation of Eq.~\ref{eq:energydensity}.
 Assuming that the density has Gaussian profile, as observed in the experiments and used to find the energy, we obtain the trap-averaged entropy per particle in terms of the reduced temperature at the trap center,
 \begin{equation}
 S=k_B\left[4-\ln\left(\frac{4}{3\sqrt{\pi}}\right)+\frac{3}{2}\ln\theta_0+\sqrt{\frac{2}{\pi}}\frac{B_2}{3}\,\theta_0^{-3/2}\right].
 \label{eq:entropy}
 \end{equation}
Then we calculate $E$ versus $\theta_0$ (using Eqs.~\ref{eq:energy},~\ref{eq:theta},~and~\ref{eq:temp}) to obtain $E$ versus $S$.  Fig.~\ref{fig:entropy} shows the average entropy per particle over the measured energy range.

\begin{figure}
\begin{center}\
\includegraphics[width=120mm]{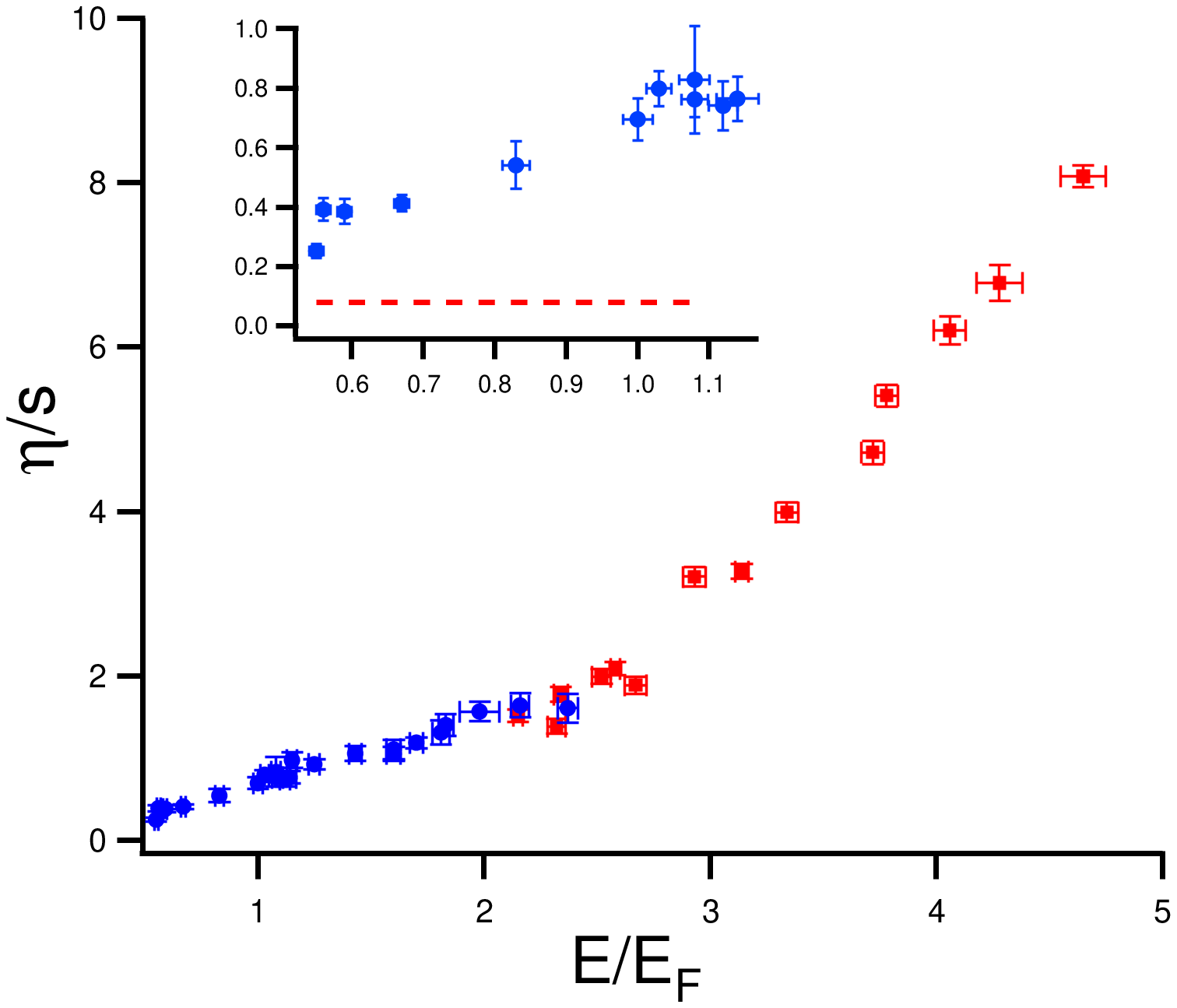}
\end{center}
\caption{Estimated ratio of the shear viscosity to the entropy density. Blue circles: Breathing mode measurements; Red squares: Anisotropic expansion measurements; Inset: Red dashed line denotes the string theory limit.  Bars denote statistical error arising from the uncertainty in $E$, $\bar{\alpha}$, and $S$~\cite{CaoViscosity}.\label{fig:etaovers}}
\end{figure}

  Fig.~\ref{fig:etaovers} shows the estimated ratio  $\eta/s$. The inset shows the low temperature behavior, which is about $5$ times the string theory limit (red dashed line) near the critical energy~\cite{ThermoLuo} $E_c/E_F = 0.7-0.8$. This result is in good agreement with recent predictions~\cite{LeClairEtaS}, where $\eta/s=4.7$. We note also that the apparent decrease of the $\eta/s$ ratio as the energy approaches the ground state $0.48\,E_F$~\cite{ThermoLuo} does not require that the local ratio  $\rightarrow 0$ as $T\rightarrow 0$, since contributions from the cloud edges significantly increase $S$  compared to the local $s$ at the center.

\section{Summary}

We have measured the trap-averaged shear viscosity coefficient as a  function of  reduced temperature at the trap center, from nearly the ground state to the two-body unitary regime. From our measurements of global entropy and energy, we have calibrated the temperature versus energy and find very good agreement with integrated local measurements as well as predictions. Using the temperature calibration, we determine the trap-averaged shear viscosity coefficient as a function of reduced temperature at the trap center, which is compared to recent predictions. We show that the best fit to our experimental expansion data is obtained with a vanishing bulk viscosity. The measured trap-averaged entropy per particle and shear viscosity are used to estimate the ratio of the shear viscosity to the entropy density. Near the transition point, the ratio is found to be $\simeq 5$ times that of a perfect fluid. More work is needed to understand the behavior of the viscosity in the low temperature regime, where the breathing mode damping rate and hence the viscosity appear to approach $0$ as $T\rightarrow 0$.

\ack This research is supported by the Physics Divisions of the
 National Science Foundation, the Army Research Office, the Air Force Office of Sponsored Research, and the
Division of Materials Science and Engineering,  the
Office of Basic Energy Sciences, Office of Science, U.S.
Department of Energy.  J.~E.~T. thanks the ExtreMe Matter Institute (EMMI) for hospitality.

\section*{References}


\begin{thebibliography}{10}
\bibitem{p-wave}
The experiments are performed far from p-wave Feshbach resonances. The relevant
  threshold energy for p-wave scattering is then comparable to the barrier
  height. Using the known $C_6$ coefficients, the barrier height for $^{40}$K
  is $280\,\mu$K, while for $^6$Li, it is 8 mK. Hence, for temperatures in the
  $\mu$K range, as used in the experiments, p-wave scattering is negligible and
  s-wave scattering dominates.
\end{thebibliography}

\end{document}